\documentclass[aps,pra,twocolumn,showpacs,amsmath,amsmath,amssymb,amsfonts,superscriptaddress]{revtex4-1}
\usepackage{graphicx}
\usepackage{dcolumn}
\usepackage{bm}
\usepackage{color}
\usepackage{multirow}

\begin{document}
\title{Controlling the dynamical scale factor in a trapped atom Sagnac Interferometer}
\author{Yijia Zhou}
\affiliation{School of Physics and Astronomy, University of Nottingham, Nottingham, NG7 2RD, UK}
\author{Igor Lesanovsky}
\affiliation{School of Physics and Astronomy, University of Nottingham, Nottingham, NG7 2RD, UK}
\affiliation{Centre for the Theoretical Physics and Mathematics of Quantum Non-equilibrium Systems, The University of Nottingham, Nottingham, NG7 2RD, United Kingdom}
\affiliation{Institut f\"ur Theoretische Physik, Universit\"at T\"ubingen, 72076 T\"ubingen, Germany}
\author{Thomas Fernholz}
\affiliation{School of Physics and Astronomy, University of Nottingham, Nottingham, NG7 2RD, UK}
\author{Weibin Li}
\affiliation{School of Physics and Astronomy, University of Nottingham, Nottingham, NG7 2RD, UK}
\affiliation{Centre for the Theoretical Physics and Mathematics of Quantum Non-equilibrium Systems, The University of Nottingham, Nottingham, NG7 2RD, United Kingdom}
\begin{abstract}
Sagnac interferometers with massive particles promise unique advantages in achieving high precision measurements of rotation rates over their optical counterparts. Recent proposals and experiments are exploring non-ballistic Sagnac interferometers where trapped atoms are transported along a closed path. This is achieved by using superpositions of internal quantum states and their control with state-dependent potentials. We address emergent questions regarding the dynamical behavior of Bose-Einstein condensates in such an interferometer and its impact on rotation sensitivity. We investigate complex dependencies on atomic interactions as well as trap geometries, rotation rates, and speed of operation. We find that temporal transport profiles obtained from a simple optimization strategy for non-interacting particles remain surprisingly robust also in the presence of interactions over a large range of realistic parameters. High sensitivities can be achieved for short interrogation times far from the adiabatic regime. This highlights a route to building fast and robust guided ring Sagnac interferometers with fully trapped atoms.
\end{abstract}
\date{\today}
\maketitle

\section{Introduction} 
\label{sec:Introduction}
Atom interferometry~\cite{Cronin2009} for precision measurements and quantum sensing~\cite{Degen2017} has become a powerful tool with applications ranging from fundamental physics~\cite{Parker2018} to absolute gravimetry~\cite{Bidel2018} and inertial sensing~\cite{Canuel2006, Fang2016}. If the effect to be measured depends on length or inertial and gravitational forces, the scaling of sensitivity with particle mass in an atom interferometer can be directly compared to its optical counterpart, promising signal gain by orders of magnitude~\cite{Clauser1988}. In a Sagnac interferometer, the resulting phase $\phi_S=2\frac{m}{\hbar}A\omega_S$ can be used to measure rotation frequency $\omega_S$. This phase is proportional to the (equivalent) mass $m$ and the area $A$ enclosed by the interferometer, factors which combine into the scale factor $\partial\phi_s / \partial\omega_s$. Despite much smaller particle flux and enclosed interferometer area, atom interferometric gyroscopes rival commercial fibre-optic devices. Sensitivities below $10^{-9}~\mathrm{rad}/\sqrt{\mathrm{s}}$ \cite{Durfee2006}, \cite{Gustavson2000} with thermal beams and stability below $10^{-9}~\mathrm{rad}/\mathrm{s}$ \cite{Dutta2016, Savoie2018} with free-falling, laser-cooled ensembles have been demonstrated, see~\cite{Barrett2014} for a recent review. In order to reduce apparatus sizes and to gain operational independence from specific conditions of gravitation and acceleration, a range of ring-shaped atom traps and guided interferometers have been proposed and implemented with various means, geometries, and objectives~\cite{Fernholz2007, Burke2009, Japha2007, Arnold2006, Morizot2006, Sherlock2011, Eckel2014, Stevenson2015, Ryu2015, Helm2015, haine_mean-field_2016,Bell2016, Navez2016, Campbell2017,Marti2015,helm_spin-orbit-coupled_2018,haine_quantum_2018,haine_quantum_2018, West2019}. Large enclosed areas are desired, and multiple cycles~\cite{Burke2009} or, equivalently, resonator approaches~\cite{Japha2007} have been proposed, although the scaling of decoherence due to longer path lengths may limit the possible benefit over physically large areas~\cite{Arnold2006}.

\begin{figure} [htbp]
\includegraphics[width=0.95\linewidth]{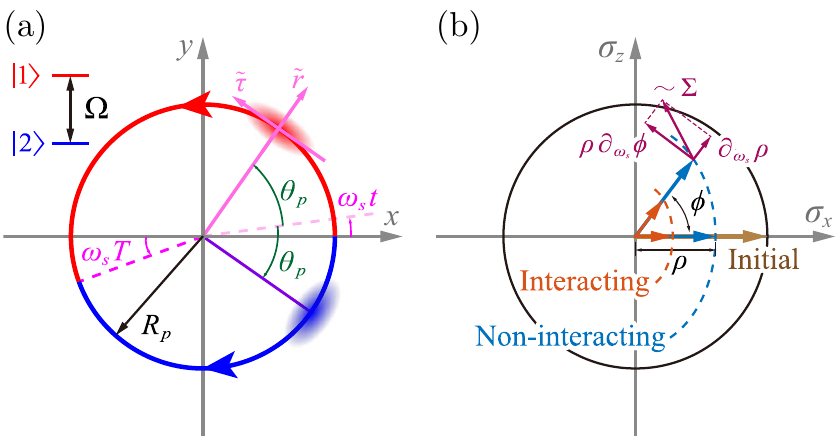}
\caption{\textbf{External and internal dynamics of the Sagnac interferometer.} (a) Atoms in a coherent superposition of two internal states $\{\left|1\right\rangle, \left|2\right\rangle\}$ (coupled with Rabi frequency $\Omega$) are initially located on the $x$-axis and transported along opposite paths. In the depicted inertial frame, the external, anti-clockwise rotation at $\omega_s$ forces the $|1\rangle$-component (red, top path) to travel a longer distance than the $|2\rangle$-component (blue, bottom path). This induces a Sagnac phase between the two states. (b) Bloch vector representation of collective, internal states and interferometer sensitivity. State vectors are shown in the $\sigma_x-\sigma_z$ plane. Referenced to the initial state (brown arrow), the final states (blue and red arrows) acquire a phase $\phi$ and show reduced contrast $\rho$. Those two quantities depend on the external rotation $\omega_s$ (dashed curves), dynamical factors, and interactions. The total change of the Bloch vector with respect to $\omega_s$ determines the interferometer sensitivity {via the dynamical scale factor}.}	\label{fig:geometry}
\end{figure}
The majority of approaches relies on the ballistic motion of particles along a closed path, but the rotation-dependent Sagnac phase can equally be measured with fully trapped atoms, i.e.\ confined in three dimensions~\cite{Stevenson2015, Navez2016, Che2018}. In such a trapped Sagnac interferometer (TSI) atomic motion is actively controlled. The necessary beam splitting and recombination arise from coherent internal state operations in conjunction with state-dependent potentials. Fully trapped atoms promise some important advantages. Interference can be observed without a standing wave phase pattern, which may require high imaging resolution~\cite{Bell2016} and interferometric stability with respect to a reference that is external to the trap, e.g.\ camera position or a standing wave light field. Particles can be accelerated to high speeds on path enclosing large areas within short times, a goal pursued by experiments on large momentum beam splitters~\cite{Mueller2008, Chiow2011, Jaffe2018}. Atomic wave packets do not disperse and their transport can be well controlled against gravity and external acceleration, where, in contrast, ballistic operation will affect the cycle time and may even preclude the enclosure of a large physical area.

Both guided and trapped interferometers have not yet reached maturity, and some intrinsic effects received little attention so far.
The discussion in the context of quantum sensing is often focused on a quantum advantage that only affects shot-to-shot noise.  But the advantage can easily become unsubstantial as many performance parameters must be considered that can be coarsely categorized as affecting precision and accuracy.  E.g., only the combination of shot-to-shot phase noise, measurement bandwidth, and scale factor leads to meaningful short-term sensitivity, and these factors may be traded against each other. Uncertainties in the scale factor affect accuracy and stability. The intrinsic effects of guided matter-wave interferometers include excitation of higher trap modes by internal and external forces, such as centripetal forces and imperfections that alter particle trajectories~\cite{Ryu2015, West2019}, potential corrugations, external acceleration, and vibration. These will affect timing, enclosed area, and interferometer contrast, and understanding their impact is complicated further by atomic interactions, quantum degeneracy and dimensionality of the atomic ensemble. These effects alter the proportionality between the measured signal and rotation, i.e.\ the \emph{signal's scale factor}, which is not simply given by the static factors that enter the expression for the Sagnac phase. An actual measurement is rather determined by a more involved, dynamical dependence of the interferometer output on external rotation, which defines a dynamical scale factor.

In this work, we investigate the dynamical scale factor in conjunction with measurement bandwidth for the case of a trapped two-mode Bose-Einstein condensate (BEC) and analyze a simple optimization scheme to achieve robust sensitivities, focusing on the slow rotation regime. Maximal sensitivities can be obtained when the spatial wave functions in the two interferometer arms remain identical. However, the transport will excite opposing center-of-mass motions in the trap, and these are not necessarily anti-symmetric between the two arms due to external rotation and atomic interactions. The path-dependent excitation severely reduces interferometer sensitivities. Through optimizing time-dependent driving profiles of the transport potential, we can robustly achieve near-maximal sensitivities at short interrogation times (hence large bandwidth) regardless of atomic interactions.

\section{Interferometer Model} 
\label{sec:Mean field model}
We consider an ensemble of $N$ atoms with two internal states $\{\left|1\right\rangle, \left|2\right\rangle\}$, as depicted in Fig.~\ref{fig:geometry}a. These can be hyperfine levels of alkali atoms (e.g.\ Rb, Cs).  Atoms can be put into coherent superposition of internal (clock) states and transported in opposite directions along a ring by state-dependent traps~\cite{Fernholz2007, Navez2016, Bentine2017}, which are guided along a ring with radius $R_p$, as shown in Fig.~\ref{fig:geometry}a. For simplicity, we assume strong confinement in the direction perpendicular to the ring ($z$-axis) assuming that the system remains in the ground state in this direction. The dynamics of the BEC are governed by two coupled, two-dimensional (2D) Gross-Pitaevskii equations (GPEs)~\cite{Williams2000},
\begin{equation}
\label{eq:gpe}
 i\hbar\frac{\partial}{\partial{t}} \psi_{j}(\mathbf{r},t) = \left(h_j  + g_{jk} \left|\psi_{k}(\mathbf{r},t)\right|^2 \right) \psi_{j} + \frac{\Omega_{jk}}{2} \psi_k(\mathbf{r},t),
\end{equation}
where $j,k=1,2$  label the two internal states (components). We have used $\Omega_{jk}$ to denote the {pulse-driven} coupling strength between the two states, with $\Omega_{jk}=0$ if $j=k$. The order parameter $\psi_j(\mathbf{r},t)$ is normalized according to $n_1+n_2=1$ with the $j$-th component occupation probability $n_j=\int|\psi_j(\mathbf{r},0)|^2\textrm{d}r$. The $j$-th component Hamiltonian is given by
\begin{equation}
 h_j=-\frac{\hbar^2}{2m}\nabla_j^2 + \frac{m\omega_r^2}{2}\tilde{r}_{j}^2 + \frac{m\omega_\tau^2}{2}\tilde{\tau}_j^2+ g_{jj} \left|\psi_{j}(\mathbf{r},t)\right|^2 ,
\end{equation}
where $m$ is the atomic mass and $\omega_r$ ($\omega_\tau$) are the trapping frequencies in the radial (azimuthal) directions. In Eq.~(\ref{eq:gpe}), we have defined local coordinate vectors $\tilde{r}_j = \left( x-R_p\cos\Theta_{j} \right)\cos\Theta_{j} - \left( y-R_p\sin\Theta_{j} \right)\sin\Theta_{j}$ and $\tilde{\tau}_j = \left( x-R_p\cos\Theta_{j} \right)\sin\Theta_{j} + \left( y-R_p\sin\Theta_{j} \right)\cos\Theta_{j}$, with respect to the trap centers at
$\Theta_{j}(t) = \pm\theta_p(t) + \omega_s t$. The trap centers are determined by the \emph{driving function} $\theta_p(t)$ and the external rotation of angular frequency $\omega_s$ that is to be measured. The boundary conditions for the driving function are $\theta_p(0)=0$ and $\theta_p(T)=\pi$. The coefficients $g_{jk}=2\sqrt{2\pi}N\hbar^2a_{jk}/(ml_z)$ quantify the strengths of intra-state ($j=k$) and inter-state ($j\neq k$) interactions, which depend on the number of particles $N $, and the effective $s$-wave scattering lengths $a_{jk}$ under the out-of-plane confinement length $l_z$~\cite{Salasnich2002}. For convenience, we scale time, energy and length  according to $t_s=1/\omega_r$, $E_s=\hbar\omega_r$ and $l_s=\sqrt{\hbar/2m\omega_r}$ in the following unless stated {explicitly}.
\begin{figure} [htb]\includegraphics[width=\linewidth]{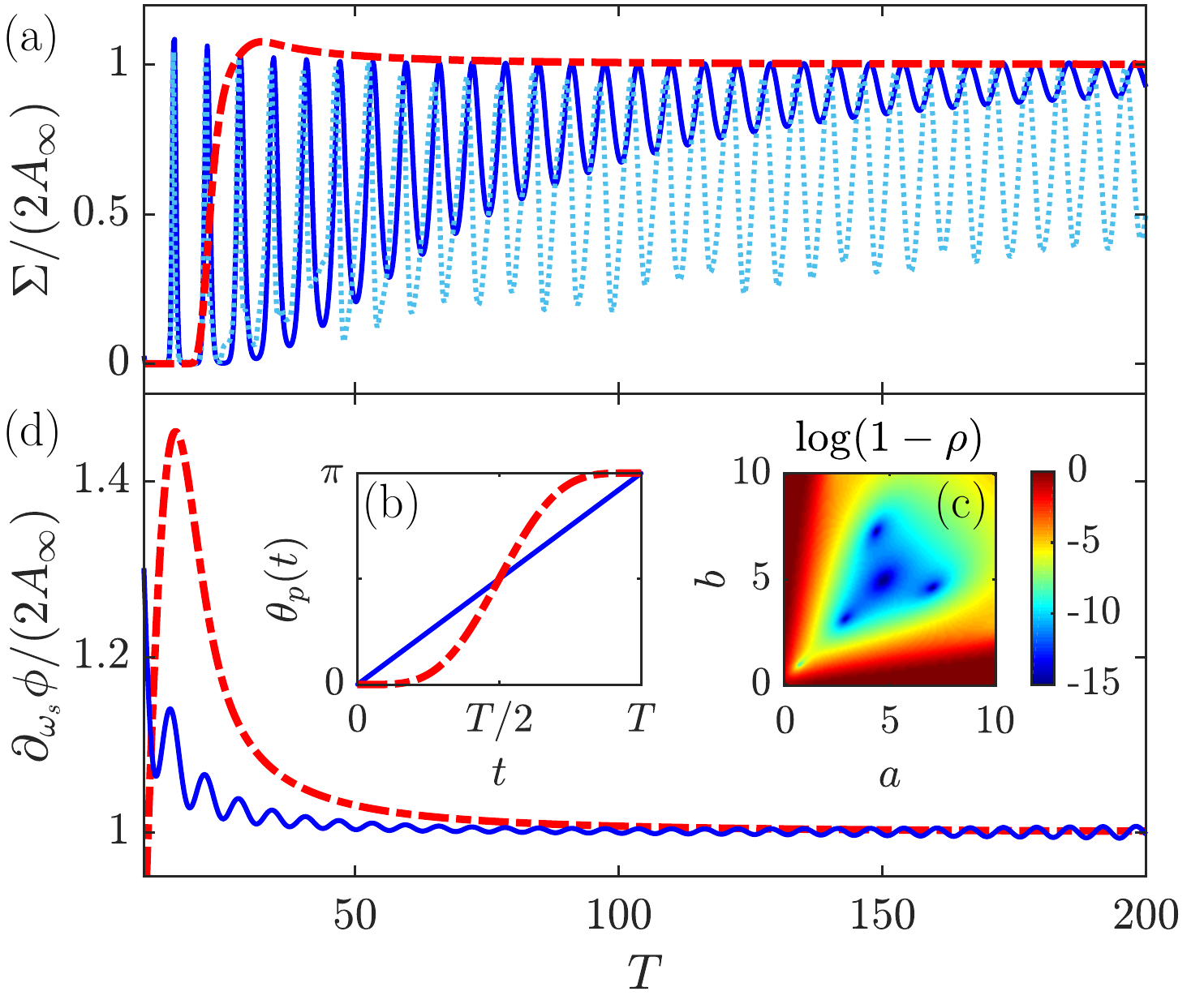}
\caption{\textbf{{Dynamical scale factor} and optimized driving function.} (a) {Dynamical scale factor} as a function of $T$ for linear driving without (solid) and with interaction $g_{11}=g_{22}=g_{12}=30$ (dotted). When applying an optimized driving function with parameters $a=b=4$ [see text for and panel (b) details], the oscillations are suppressed and the {scale factor} approaches the static value for $T>30$ (dashed). (b) Driving function for constant speed ($a=b=0$, solid), and optimal driving with $a=b=4$ (dashed). (c) Interferometer contrast $\rho$ as a function of the parameters $a$ and $b$ of the profile (\ref{trial}) for an interrogation time $T=50$. (d) Behavior of the rotation dependent phase factor of the interferometer without (solid) and with (dashed) optimization.}
\label{fig:optimal}
\end{figure}

\section{Sensitivity of the Sagnac interferometer} 
\label{sec:Sensitivity} To operate the TSI, we first create a coherent superposition state $(|1\rangle+|2\rangle)/\sqrt{2}$ by performing a fast $\pi/2$-pulse on a BEC in the internal state $|1\rangle$. Here and in the following, we neglect dynamics of the spin rotation process as $\Omega_{12}$ ($\Omega_{21}$) is on the order of MHz, which is far larger than typical energy scale (kHz) of the BECs.  To take account of atomic interactions, the ground state and dynamics of the trapped BECs are obtained by numerically solving Eq.~(\ref{eq:gpe}) with an imaginary and real-time algorithm, respectively.  The two states are then guided along a ring in opposite directions (see Fig.~\ref{fig:geometry}a). When they are recombined after the interrogation time $T$ ~\cite{Barrett2014}, a second fast $\pi/2$-pulse is applied to convert the accumulated phase difference into a population difference between the two states. The rotation frequency $\omega_s$ is then encoded in the expectation value $\left\langle \sigma_z \right\rangle=n_{2}-n_{1}$. 
\begin{figure} [htbp]\includegraphics[width=\linewidth]{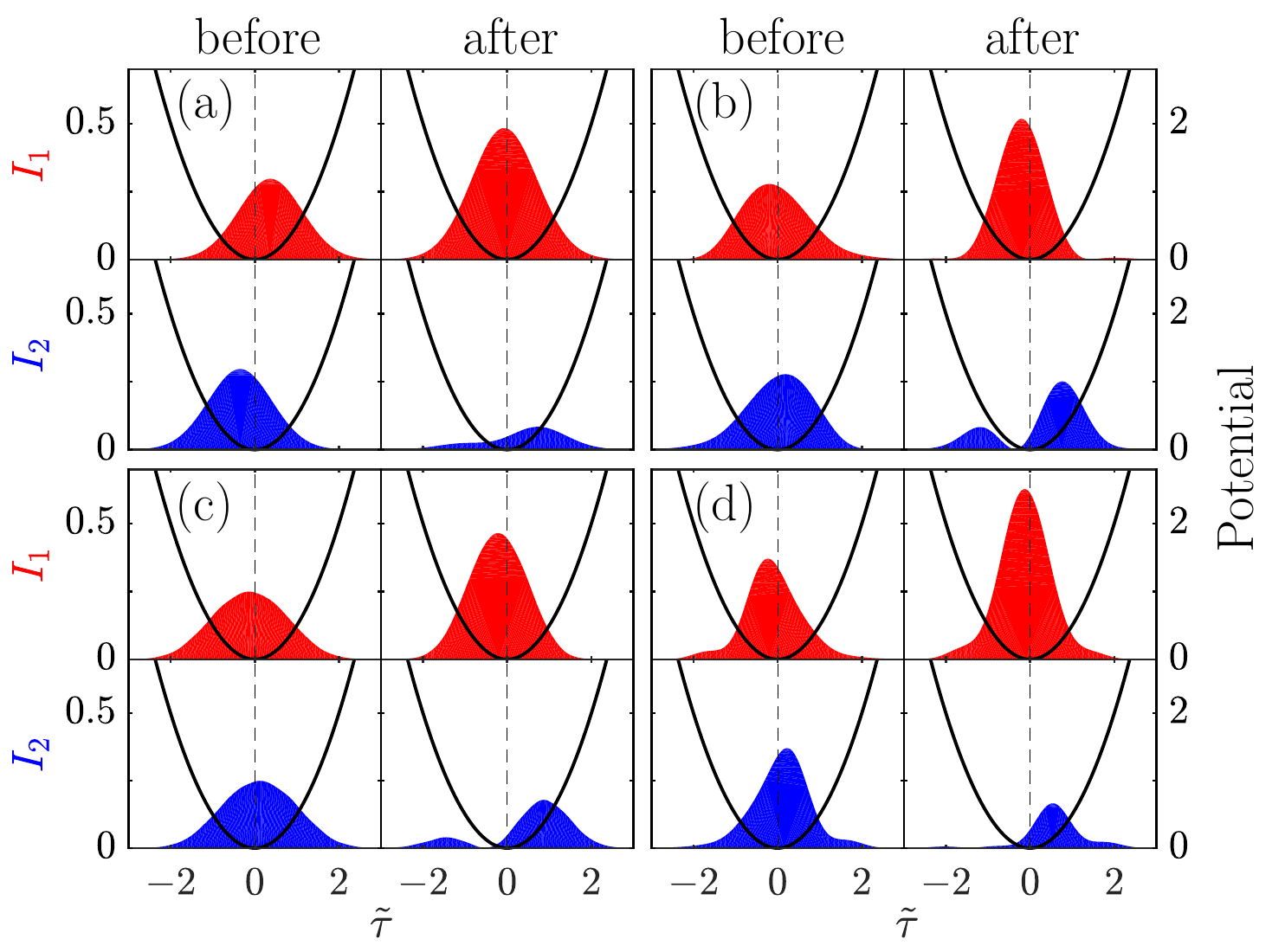}
	\caption{\textbf{Internal BEC dynamics.} Spatial densities of the BEC components projected on the $\tilde{\tau}$ direction. Density snapshots of the states $|1\rangle$ (upper row) and $|2\rangle$ (lower row) immediately before and after the second (recombination) $\pi/2$-pulse are shown in each panel. Panels correspond to different interaction strengths: (a) $g_{11}=g_{22}=g_{12}=0$, (b) $g_{11}=g_{22}=g_{12}=30$, (c) $g_{11}=g_{22}=30, g_{12}=0$, and (d) $g_{11}=g_{22}=0, g_{12}=30$.  The parameters used in the calculations are $\omega_r=\omega_{\tau}=1$, $T=100$. The trapping potential (solid) and trap center (dashed) along the $\tilde{\tau}$ axis (see main text for details) are shown.}\label{fig:ProjectedWavefunction}
\end{figure}

The detection can be sensitive, if the differential response of population difference to rotation rate is large. At the same time, noise should be low and the interrogation time $T$ should be short, equivalent to high measurement bandwidth ($\sim1/T$). Combining these considerations, the figure of merit is given by the interferometer's short term sensitivity $S$, which equals angular random walk of the time integrated rotation estimate. Assuming shot noise of uncorrelated particles near balanced output, it is given by 
\begin{equation}
S=\left[\sqrt{\dot{N}}{\Sigma(\omega_s)}\right]^{-1},
\end{equation}
which depends on the dynamical scale factor $\Sigma(\omega_s)$ and improves for increased particle throughput $\dot{N}=N/T$, i.e.\ the number of particles per cycle $N$ and the cycle time $T$. For correlated particles, this can be improved to the Heisenberg limit $S_\mathrm{HL}=\left[N{\Sigma(\omega_s)}\right]^{-1}\sqrt{T}$~\cite{Ragole_interaction}.
The dynamical scale factor can be expressed as
\begin{equation} \label{sensitivity}
 \Sigma(\omega_s) = \frac{ \partial\left\langle \sigma_z \right\rangle}{\partial \omega_s}=\frac{\langle\sigma_z\rangle}{\rho}\frac{\partial \rho}{\partial \omega_s} + \rho\cos\phi\frac{\partial \phi}{\partial \omega_s},
\end{equation}
which measures changes of the expected signal $\left\langle \sigma_z \right\rangle$ with respect to angular rotation frequency $\omega_s$. Here, we used the parameterization $\left\langle \sigma_z \right\rangle = \rho\sin\phi$, where $\rho$ and $\phi$ describe length (contrast) and orientation (phase) of the Bloch vector, see Fig.~\ref{fig:geometry}b. Both quantities are determined by the spatial overlap of the two states $\int \psi_2^*(\mathbf{r},T)\psi_1(\mathbf{r},T)\textrm{d}\mathbf{r} = \rho e^{i\phi}$~\cite{Stevenson2015} (see Appendix A for details). Imperfect spatial overlap occurs if the center-of-mass positions or momenta of the two states do not coincide at the end of the sequence even if their wave functions share similar shapes~\cite{roura_overcoming_2014,japha_general_2019}. More importantly, atomic interactions and external rotation together can significantly distort wave functions of the two states asymmetrically, which significantly reduces the overlap. In the following, we will focus on influences of the latter on the sensitivity.

In the slow rotation limit $\omega_s\to 0$, the derivative $\partial\rho/\partial\omega_s$ of the scale factor Eq.~(\ref{sensitivity}) vanishes because $\rho$ must be an even function of $\omega_s$. Here, the maximum scale factor (obtained by setting the phase reference such that $\phi=0$ at $\omega_s=0$) reduces to
 \begin{equation}
 \label{eq:redsensitivity}
 \tilde{\Sigma} =\Sigma(\omega_s=0) = \left. \rho\frac{\partial\phi}{\partial\omega_s} \right|_{\omega_s=0},
 \end{equation}
which solely depends on $\rho$ and the phase gradient $\partial\phi/\partial \omega_s$. In the following we will investigate how these two parameters depend on the atomic interaction, the trap aspect ratio, and the interrogation time $T$.

First, we consider a simple linear driving profile $\theta_p(t)=t\pi/T$ and a non-interacting BEC. Fig.~\ref{fig:optimal}a shows the scale factor obtained by numerically solving the coupled GPEs with a small rotation (we take $\omega_s=10^{-3}\omega_r$ and $R_p=10$, throughout the work). The scale factor oscillates as a function of the interrogation time $T$, with decreasing amplitude for increasing $T$~\cite{Stevenson2015}. The reason is that in-trap oscillations excited by the initial acceleration will be stopped by the final deceleration, if the driving profile does not contain the corresponding spectral components~\cite{Stevenson2015}. The oscillation frequency is thus the same as the trapping frequency $\omega_r$, which signifies the elementary excitations in the moving trap that affects the interference of the two matter-waves. In the limit $T\rightarrow\infty$, the scale factor is approximately given by
\begin{equation} \label{eq:AdSensitivity}
\tilde{\Sigma}_{\infty} \approx \frac{2A_{\infty}}{ [1-\left(\pi/T\right)^2]^2}.
\end{equation}
It thus approaches the conventional scale factor $2A_\infty$, where $A_\infty=\pi R_p^2$ is the area of the ring (in scaled units of $l_s^2$). The denominator accounts for the correction of the centrifugal effect (see Appendix C). This dependence provides a first example for a \emph{dynamical} scale factor: the centrifugal forces lead to an increase in the area enclosed by the atomic trajectories. This effect is present in higher dimensions (2D and 3D) whilst absent in 1D models.

Scale factors change qualitatively when inter- and intra-state interactions are taken into account. We observe that the oscillations of the dynamical scale factor increase drastically at intermediate interrogation times, and their amplitudes decrease much slower with increasing $T$ (Fig.~\ref{fig:optimal}a) than in the non-interacting case. The reduction of scale factors arises from the fact that collective modes of the BEC are excited when atoms are transported non-adiabatically~\cite{Pethick2008} around the ring. To illustrate this, we evaluate the projected BEC densities on the $\tilde{\tau}$ axis, $I_j=\int d\tilde{r}|\psi_j|^2$ right before and right after the second $\pi/2$-pulse. As shown in Fig.~\ref{fig:ProjectedWavefunction}a for a non-interacting BEC, the densities of the individual components may be shifted oppositely from the trap center before the second pulse, leading to incomplete conversion of phase into population difference. In addition, the density profiles change with increasing interaction strengths (Fig.~\ref{fig:ProjectedWavefunction}b-d). When intra-state interactions dominate (Fig.~\ref{fig:ProjectedWavefunction}d), the wave packets distort significantly from a Gaussian shape during the transport.

\section{Optimization for a non-interacting and interacting BEC} 
\label{sec:Optimization non-interacting}
In the following we aim to avoid the path dependent excitation of the BEC components in order to reach maximal scale factor $\tilde{\Sigma}_{\infty}$ as well as short interrogation times.  Previous studies~\cite{Stevenson2015, Che2018} have considered ideal driving functions of a non-interacting BEC by excluding frequency components at the trapping frequency, $\int_0^T \sin[\theta_p(\tau)]e^{i\tau}\textrm{d}\tau=0$. However, this condition does not avoid oscillations during the transport and it is insufficient when interactions are non-negligible.

During the transport, the BEC should be accelerated (decelerated) slowly at $t\rightarrow0$ ($t\rightarrow T$) to avoid dynamical excitations, which is satisfied by the nonlinear driving function
\begin{equation}
\label{trial}
\theta_p(t) =\frac{\pi}{B(a+1,b+1)} \int_0^t \left(\frac{t'}{T}\right)^a \left(1-\frac{t'}{T}\right)^b \textrm{d}t',
\end{equation}
where the (Beta-)function $B(a,b)$ ensures normalization, to meet the boundary condition $\theta_p(T) = \pi$. This driving function is a convenient choice and has been applied to other optimization problems, e.g., conformal antenna arrays~\cite{Boeringer2004}. It generally has a sigmoidal form and includes the linear ramp as a  limiting case, as shown in Fig.~\ref{fig:optimal}b. 

Using this driving function, the dynamical scale factor for a non-interacting BEC is shown in Fig.~\ref{fig:optimal}a as a function of the interrogation time $T$. The oscillations have vanished and the scale factor quickly approaches the static value $\tilde{\Sigma}_{\infty}$ already at short times. The improvement of sensitivity and robustness is rooted in the fact that dynamical excitations of the BEC are suppressed significantly as the driving function has a vanishing slope at the beginning and end of the interferometer operation (see Fig.~\ref{fig:optimal}b), which ensures the BEC is accelerated and decelerated slowly. Note, that this behavior is largely independent of precise choices of the parameters $a$ and $b$, which can be seen in Fig.~\ref{fig:optimal}d.

\begin{figure} [htbp]\includegraphics[width=\linewidth]{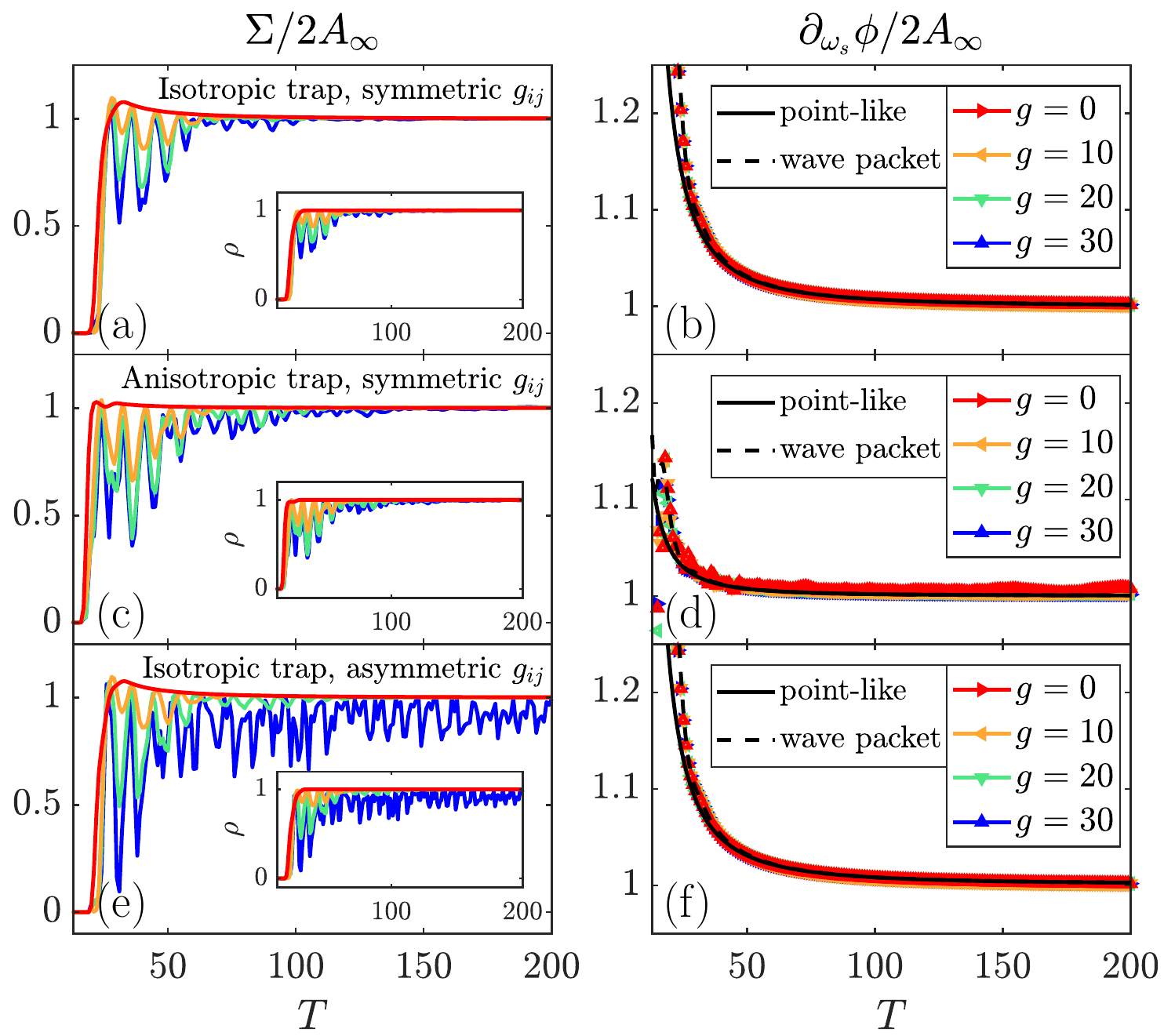} \caption{\textbf{Role of trap anisotropy and interactions.} Dynamical scale factors, normalized to the static value, (left column) and phase gradient (right column) as a function of interrogation time $T$ for the driving function parameters $a=b=4$. We consider three cases: (a,b) symmetric interactions $g_{11}=g_{22}=g_{12}=g$ and isotropic traps $\omega_r=\omega_\tau=1$, (c,d) symmetric interactions $g_{11}=g_{22}=g_{12}=g$ and anisotropic traps $\omega_r=1$, $\omega_\tau=0.5$, (e,f) asymmetric interactions $g_{11}=g_{22}=10$, $g_{12}=g$ and isotropic traps $\omega_r=\omega_\tau=1$. Insets in the left panels show the contrast $\rho$. The phase gradients are displayed together with numerical calculations of the path-enclosed area for a classical point-like particle (black solid lines) and an average over the  BEC wave packets (dashed lines).}\label{fig:isotropic}
\end{figure}

To study a TSI implemented with an interacting BEC, we first consider a scenario where atomic interactions are symmetric ($g_{11}=g_{22}=g_{12}=g$) and the trapping potentials are isotropic ($\omega_r=\omega_{\tau}$). We use the same driving function $\theta_p(t)$ with parameters $a=b=4$ as in the case of a non-interacting BEC. In Fig.~\ref{fig:isotropic}a, the resulting dynamical scale factors are presented for various interaction strengths $g$. Although oscillations of the scale factor reemerge with stronger interactions, their amplitudes quickly decrease with increasing $T$. For the strongest interactions considered in these examples ($g=30$), the dynamical scale factor settles near the maximal value $\tilde{\Sigma}_{\infty}$ for $T>100$.

Similar observations hold in the case of anisotropic trapping potentials  (Fig.~\ref{fig:isotropic}c,d) as well as for non-symmetric atomic interactions (Fig.~\ref{fig:isotropic}e,f). To illustrate effects due to trap anisotropy, we consider an example where $\omega_r=2\omega_{\tau}$ while the interactions are still symmetric. Due to the strong radial trapping, one expects a weaker centrifugal force. As a result, the maximal value of the scale factor slightly decreases at small $T$ for non-interacting BECs, as shown in (Fig.~\ref{fig:isotropic}c). In the presence of two-body interactions, dynamical scale factors oscillate with larger amplitudes in the small $T$ region. Increasing $T$, the optimized scale factor quickly approaches the maximal value. When the inter- and intra-state interactions differ,  i.e.\ for $g_{11}=g_{22}=g\neq g_{12}$, we find that robust dynamical scale factors can be obtained when $g_{12}$ is smaller or comparable to $g$.  However, the optimized (non-interacting) driving function becomes less efficient when $g_{12}$ is much greater than $g$. We attribute this to the fact that strong repulsion~\cite{Pethick2008} between the two BEC components causes immiscibility and prevents them from overlapping in space (see Fig.~\ref{fig:ProjectedWavefunction}d), leading to reduced contrast. This, however, is not a major issue in realistic experiments as the inter- and intra-state scattering lengths can be very similar (e.g.\ Rb atoms).

An interesting observation is that in all the considered cases, the scale factor is mostly influenced by a reduction of contrast rather than through the phase gradient. This can be seen by the very similar dependencies of scale factor and contrast on the interrogation time $T$, as shown in the insets to Fig.~\ref{fig:optimal}a,b,c. The scale factor exhibits a weak dependence on the phase gradient at intermediate interrogation times $20<T<50$ (see Fig.~\ref{fig:isotropic}b,d,f), where values going beyond the static scale factor are achieved due to non-negligible centrifugal forces. An important finding is that the phase gradient is largely immune to atomic interactions and trap geometry, which we attribute to the suppression of radial center-of-mass oscillations also in the presence of interactions. As shown in Fig.~\ref{fig:isotropic}b,d,f, and similar to the optimized response in Fig.~\ref{fig:optimal}d, the phase gradient decreases smoothly with increasing $T$, approaching the static scale factor for $T\rightarrow\infty$. In fact, the scale factor agrees closely with numerical calculations of the path enclosed area for a point-like classical particle  (see Appendix C for details), which is equivalent to trajectories of the center-of-mass of the BEC~\cite{japha_motion_2002,meister_chapter_2017}.  We find a good agreement even when we weigh the point-wise particle's position with the BEC wave packet. Therefore, an accurate measurement of rotation can be obtained by adaptive phase estimation protocols that co-estimate the contrast also in the case of uncertain dynamics~\cite{Wiebe2016,Lumino2018}.

Finally, we present some experimental parameters relevant to this study. Using $^{87}$Rb as an example, time-averaged-adiabatic-potential traps \cite{Navez2016} can confine the atoms in radial direction at frequency $\omega_r \approx 2\pi\times127$ Hz, and $\omega_z \approx 2\pi\times206$ Hz in $z$-direction. This results in the unit time, length and energy to be $t_s=1.26$ ms, $l_s=1.70$ $\mu$m, and $E_s=8.35\times10^{-32}$J respectively. The scattering length of $^{87}$Rb is nearly identical in $|F=1\rangle$ and $|F=2\rangle$ states for either intra-state or inter-state interactions. The difference is below 1\% and the average value is $\bar{a}_s\approx98a_0$, where $a_0$ is Bohr radius \cite{Egorov2013}. In the 2D model, the effective interaction strength is $g_{ij}=N\sqrt{m \omega_z/2\pi\hbar}(4\pi\hbar^2/m)a_{ij}$ where $N$ is the number of atoms \cite{Salasnich2002}. Using $E_s$ and $l_s$, we obtain $g=4.39\times10^{-3} N E_s l_s^2$. To change the relative interaction strength, we can, e.g., vary the number of atoms in the BEC. For example,  the dimensionless interaction strength will be $g=43.9$ for $10^4$ atoms. Tuning the interaction strength will allow for exploration of the effects predicted by this investigation.

\section{Conclusion} 
\label{sec:Conclusion}
We analyzed the dependence of the dynamical scale-factor of guided Sagnac interferometers in the slow rotation regime with respect to transport parameters, atomic interactions, and trap symmetries. Employing a simple optimized driving function for the transport of atoms in state-dependent potentials reduces path-dependent excitations and achieves maximal sensitivities at moderate interrogation times, typically tens of trap oscillation periods, for both ideal and interacting BECs. Our theoretical study is important to guide current experimental efforts on building robust and fast Sagnac interferometers with fully trapped atomic gases. It lays a foundation for further analysis of other experimentally relevant parameters, such as atom number fluctuations~\cite{quantum_noise1,rocco_fluorescence_2014}, imperfect state operations, and finite temperatures~\cite{Stevenson2015}.

\acknowledgments We thank M. Fromhold, M. Greenaway, G. Raggi, S. Wu for fruitful discussion. We are grateful for access to the University of Nottingham High Performance Computing Facility. Y. Zhou is supported by China Scholarship Council (CSC) No. 201606100035. The research leading to these results has received funding from the European Research Council under the European Union’s Seventh Framework Programme (FP/2007-2013)/ERC Grant Agreement No. 335266 (ESCQUMA) as well as from the Engineering and Physical Sciences Research Council (Grant Nr. EP/M013294/1). I.L. gratefully acknowledges funding through the Royal Society Wolfson Research Merit Award.

\newpage

\begin{appendix}
	
	\section{Derivation of the dynamical scale factor}
	The state of the two-component BEC is represented by a spinor,
	\begin{equation}
	\Psi(t) = \frac{1}{\sqrt{2}} \begin{pmatrix} \psi_1(\mathbf{r},t) \\ \psi_2(\mathbf{r},t) \end{pmatrix},
	\end{equation}
	where $\psi_1(\mathbf{r},t)$ and $\psi_2(\mathbf{r},t)$ denote the spatial wave functions corresponding to the internal states $|1\rangle$ and $|2\rangle$, respectively. Applying the coupling field with Rabi frequency $\Omega$ is equivalent with multiplying the spinor with the operator,
	\begin{equation}
	\hat{R}_\varphi(\theta_r,\phi_r) = \begin{pmatrix}
	\cos\frac{\varphi}{2} - i\sin\frac{\varphi}{2}\cos\theta_r &
	-i\sin\frac{\varphi}{2}\sin\theta_r e^{-i\phi_r} \\
	-i\sin\frac{\varphi}{2}\sin\theta_r e^{i\phi_r} &
	\cos\frac{\varphi}{2} + i\sin\frac{\varphi}{2}\cos\theta_r
	\end{pmatrix}.\nonumber
	\end{equation}
	where $\varphi=\Omega t$ is the pulse area and $(\theta_r, \phi_r)$ are reference phases. 
	
	The first $\pi/2$-pulse of the interferometer protocol creates a superposition state of $|1\rangle$ and $|2\rangle$ with equal populations, $\int|\psi_1(\mathbf{r},t=0)|^2{\rm d}\mathbf{r} = \int|\psi_2(\mathbf{r},t=0)|^2{\rm d}\mathbf{r}$.  After the interrogation time $T$ the atoms are subject to a second $\pi/2$-pulse, after which the average population difference is given by
	\begin{align}
	\langle\hat{\sigma}_z\rangle &= \left\langle \Psi(T) \middle| \hat{R}_{\pi/2}^\dag(\theta_r,\phi_r) \hat{\sigma}_z \hat{R}_{\pi/2}(\theta_r,\phi_r) \middle| \Psi(T) \right\rangle \nonumber \\
	&= \frac{1}{2} \cos^2\theta_r \int \left( |\psi_1(\mathbf{r},T)|^2 - |\psi_2(\mathbf{r},T)|^2 \right){\rm d}\mathbf{r} \nonumber \\
	&\,+ {\rm Re}\Big[\sin\theta_r e^{i\phi_r} \big( \cos\theta_r+i \big) \int \psi_1^*(\mathbf{r},T)\psi_2(\mathbf{r},T) {\rm d}\mathbf{r}  \Big] \nonumber\\
	&= \rho \sin\theta_r \Big[ \cos\theta_r\cos(\phi-\phi_r) + \sin(\phi-\phi_r) \Big].
	\end{align}
	Here we have introduced the spatial overlap $\int \psi_1(\mathbf{r},T)\psi_2^*(\mathbf{r},T){\rm d}\mathbf{r} = \rho e^{i \phi}$. 
	
	Using the fact that $\partial\rho/\partial\omega_s=0$ at $\omega_s=0$ (see main text), we find the scale factor to be
	\begin{equation}
	\frac{\partial\langle\hat\sigma_z\rangle}{\partial\omega_s} = \sin\theta_r\Big[ -\cos\theta_r\sin(\phi-\phi_r) + \cos(\phi-\phi_r) \Big] \rho\frac{\partial\phi}{\partial\omega_s},\nonumber
	\end{equation}
	whose maximum is achieved when setting the reference phases $\theta_r = \pi/2$ and $\phi_r-\phi=k\pi$ with $k$ being an integer.
	
	\section{Analytical Solutions for two dimensional BECs} \label{Append.A}
	The dynamics of two-dimensional non-interacting BECs can be solved analytically. To this end we first transform to a rotating frame with angular frequency $\omega_s$ using unitary operator $\hat{U}_{1,2}(t) = \exp\left[i\omega_st\hat{L}_z\right] = \exp\left[\omega_st\left(x\frac{\partial}{\partial{y}}-y\frac{\partial}{\partial{x}}\right)\right]$. The Hamiltonians $h_1$ and $h_2$ (with $g_{11}=g_{22}=0$), which are given in the main text, then become
	\begin{align} \label{HamiltonianInRotatingFrame}
	\hat{h}'_{1,2} &= \hat{U}_{1,2}\hat{h}_{1,2}\hat{U}^\dagger_{1,2} + i\hat{U}^\dag_{1,2}\frac{\partial}{\partial t}\hat{U}_{1,2} \nonumber \\
	&= -\frac12\left(\frac{\partial^2}{\partial{x}^2}+\frac{\partial^2}{\partial{y}^2}\right) + \frac12\left(x^2+y^2+R_p^2\right)\nonumber\\
	&\, -R_px\cos\theta_p(t) \mp R_py\sin\theta_p(t) + i\omega_s \left( x\frac{\partial}{\partial{y}} - y\frac{\partial}{\partial{x}} \right).
	\end{align}
	\begin{figure}[htbp]
	\centering
	\includegraphics[width=0.8\linewidth]{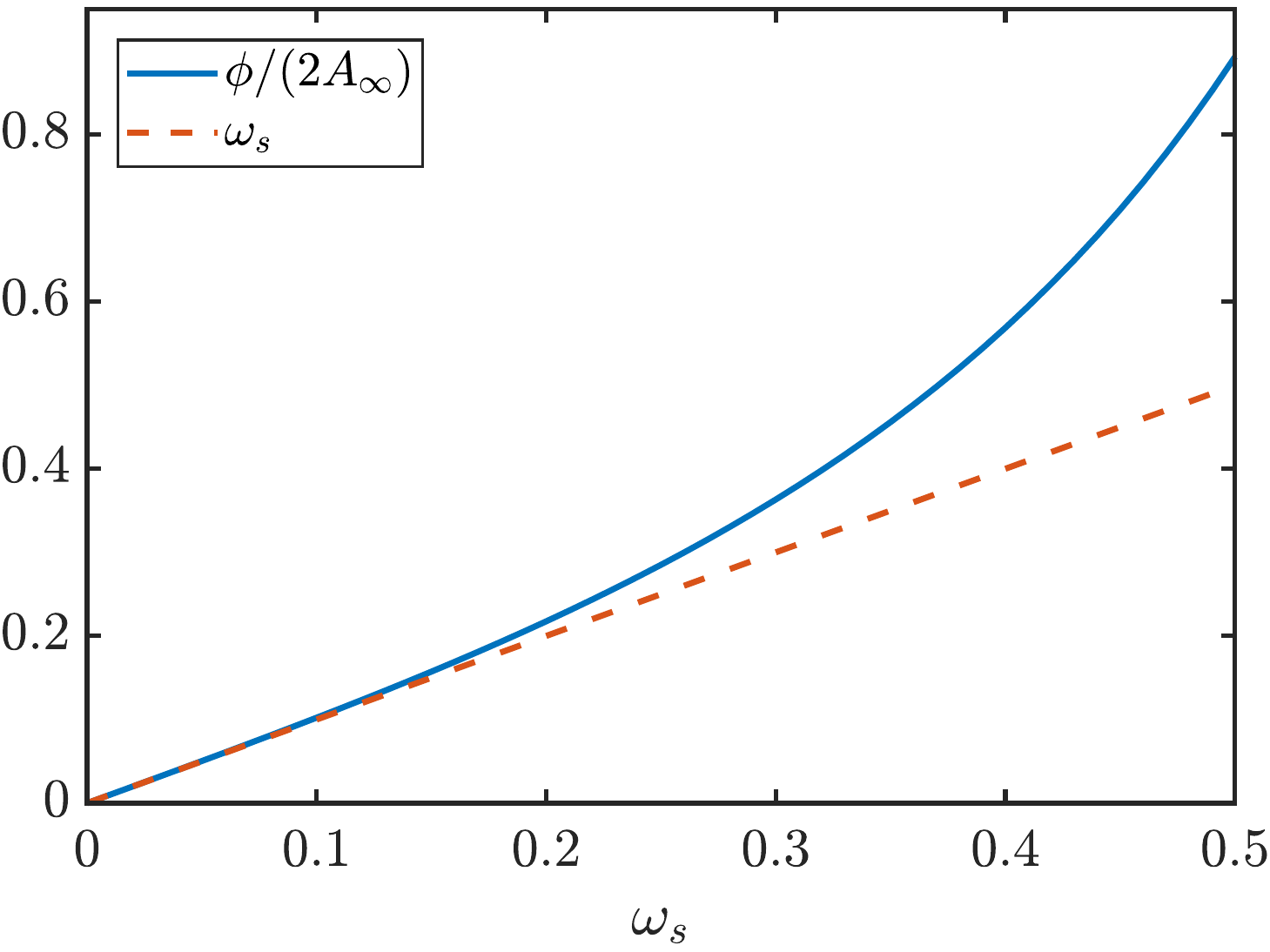}
	\caption{Interference phase $\phi$ as a function of $\omega_s$. A linear relation is found when $\omega_s<0.1$. }
	\label{fig:linearity}
\end{figure}
	
	In the next step we introduce the ladder operators
	\begin{eqnarray}
	\hat{a} &=& \frac12 \left( x+iy+\frac{\partial}{\partial{x}} + i\frac{\partial}{\partial{y}} \right)\\
	\hat{a}^\dag &=& \frac12 \left( x-iy-\frac{\partial}{\partial{x}} + i\frac{\partial}{\partial{y}} \right)\\
	\hat{b} &=& \frac12 \left( x-iy+\frac{\partial}{\partial{x}} - i\frac{\partial}{\partial{y}} \right)\\
	\hat{b}^\dag &=& \frac12 \left( x+iy-\frac{\partial}{\partial{x}} - i\frac{\partial}{\partial{y}} \right).
	\end{eqnarray}
	and the Hamiltonians \eqref{HamiltonianInRotatingFrame}, expressed in terms of these operators, acquire the following form:
	\begin{align}
	\hat{h}'_{1,2} &= \left(1+\omega_s\right)\hat{a}^\dag\hat{a} + \left(1-\omega_s\right)\hat{b}^\dag\hat{b} + 1 + \frac{R_p^2}{2}  \nonumber\\
	&\, - \frac{R_p}{2}\left(\hat{a}^\dag e^{\pm i\theta_p(t)} + \hat{a} e^{\mp i\theta_p(t)} + \hat{b}^\dag e^{\mp i\theta_p(t)} + \hat{b} e^{\pm i\theta_p(t)}\right).
	\end{align}
	Both, $h'_1$ and $h'_2$ describe two sets of uncoupled, linearly driven oscillators. Their time evolution is solved via the ansatz $|\psi_j\rangle=e^{i\phi_j}\left| \alpha_j, \beta_j \right\rangle$. Here, $\phi_j$ is a global phase and $\left| \alpha_j \right\rangle$ and $\left| \beta_j \right\rangle$ are coherent states, i.e.\ eigenstates of the operators $\hat{a}$ and  $\hat{b}$, respectively. The dynamical evolution of the coherent state amplitudes and the phase is governed by the following equations:
	\begin{subequations}
		\begin{align}
		\frac{\textrm{d}}{\textrm{d} t}\alpha_{1,2} &= -i\left(1+\omega_s\right)\alpha_{1,2} + i\frac{R_p}{2}e^{\pm i\theta_p}, \\
		\frac{\textrm{d}}{\textrm{d} t}\beta_{1,2} &= -i\left(1-\omega_s\right)\beta_{1,2} + i\frac{R_p}{2}e^{\mp i\theta_p}, \\
		\frac{\textrm{d}}{\textrm{d} t}\phi_{1,2} &= \frac{R_p}{2}\textrm{Re}\left[\alpha_{1,2}e^{\mp i\theta_p} + \beta_{1,2}e^{\pm i\theta_p} \right].
		\end{align}
	\end{subequations}
	By directly integrating these equations, we find the solutions,
	\begin{subequations} \label{AnalyticalSolutions}
		\begin{align}
		\alpha_{1,2}(t) &= \frac{R_p}{2(1+\omega_s)} \left[1 - i \int_0^t \dot{\theta}_p(t') e^{ i(1+\omega_s)(t'-t)\pm i\theta_p(t') } \textrm{d}t' \right], \\
		\beta_{1,2}(t) &= \frac{R_p}{2(1-\omega_s)}  \left[1 + i \int_0^t \dot{\theta}_p(t') e^{ i(1-\omega_s)(t'-t)\mp i\theta_p(t') } \textrm{d}t' \right],  \\
		\phi_{1,2}(t)  &= \frac{R_p}{2} \textrm{Re}\left[ \int_0^t \alpha_{1,2}(t') e^{\mp i \theta_p(t')} + \beta_{1,2}(t') e^{\pm i \theta_p(t')} \textrm{d}t' \right].
		\end{align}
	\end{subequations}
	Here, we have assumed that the oscillators are initially in their ground states and that $\theta_p(0)=0$. 
	\begin{figure} [htbp]
	\centering
	\includegraphics[width=0.8\linewidth]{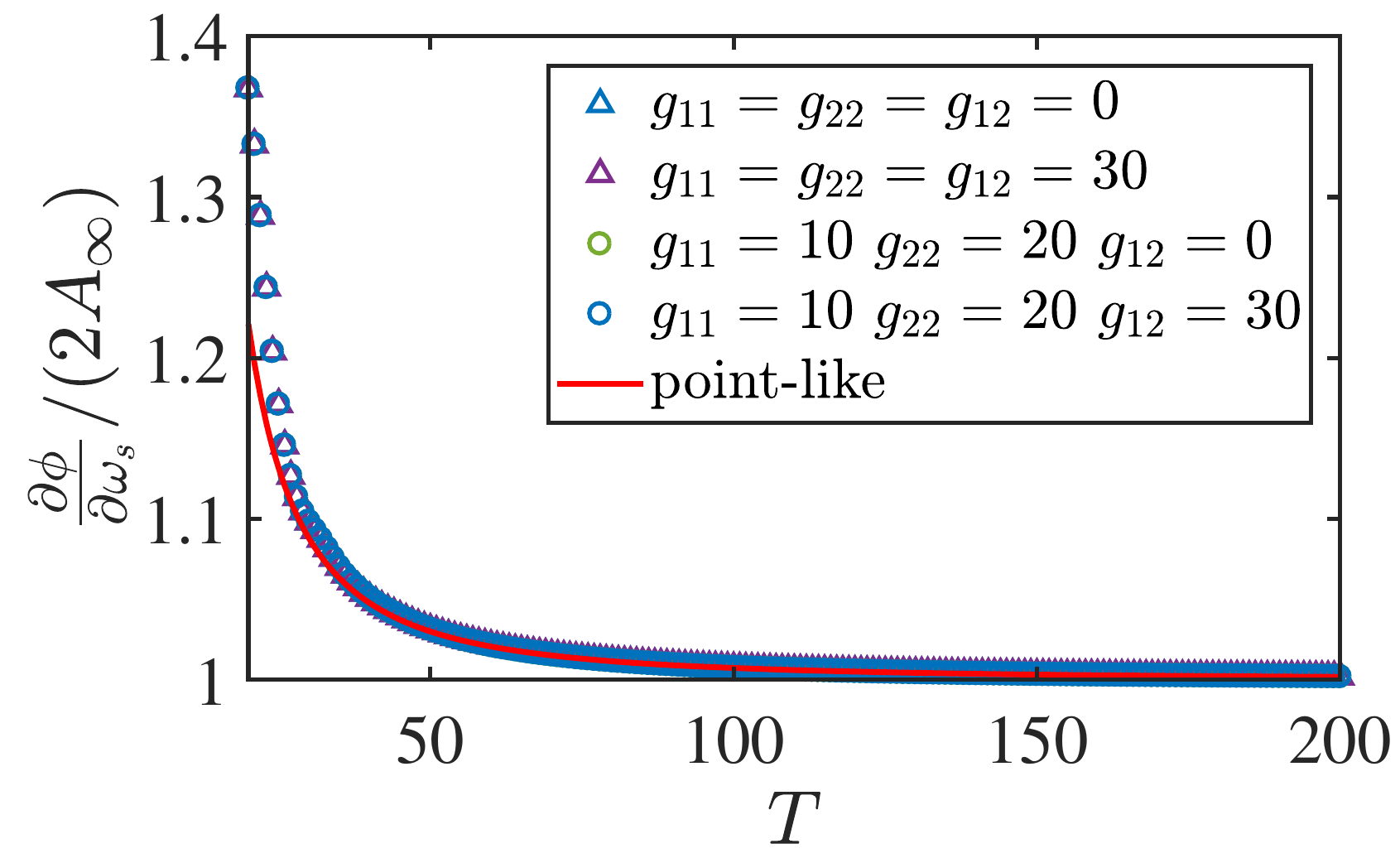}
	\caption{Phase gradient as a function of the interrogation time $T$. The estimate based on a point-like particle subject to the centripetal force agrees well with numerical calculations and is largely independent from the presence or absence of interactions.}
	\label{fig:phasestability}
\end{figure}

	For a given driving profile $\theta_p(t)$ the above expressions can be evaluated either analytically (in special cases) or numerically. The result then allows us to calculate quantities such as the time-dependent spatial overlap between the two internal states,
	\begin{align}
	&\int \psi_1(\mathbf{r})\psi_2^*(\mathbf{r})\textrm{d}\mathbf{r} \nonumber\\
	&= e^{i (\phi_1-\phi_2 ) } \left\langle \alpha_2 \mid \alpha_1 \right\rangle \left\langle \beta_2 \mid \beta_1 \right\rangle \nonumber \\
	&= e^{i \big[ \phi_1-\phi_2 + \textrm{Im} (\alpha_2^*\alpha_1 + \beta_2^*\beta_1 ) \big]} e^{-\frac12(|\alpha_1-\alpha_2|^2 + |\beta_1-\beta_2|^2)}.
	\end{align}

The solution furthermore allows is to calculate the dependence of the interference phase $\phi$ on the rotation angular velocity $\omega_s$, as is shown in Fig.~\ref{fig:linearity}. The discussion in the main text focusses on the linear regime which is achieved when $\omega_s<0.1$.
	\section{Classical estimate for the interference phase}
	A ``classical" estimate for the interference phase can be obtained by considering the motion of a point-particle (initially located at the trap center) along the ring. When rotating along the ring, the particle experiences a centripetal force (due to a finite angular momentum $R_p\dot{\theta}_p$) which will dynamically alter the radius $R_p$, i.e.\ the particle's displacement from the ring center. Balancing the centripetal and trapping forces (in the scaled units),
	\begin{equation}
	\dot{\theta}_p^2 R = (R-R_p).
	\end{equation}
	we obtain the new radius to be $R=R_p/(1-\dot{\theta}_p^2)$. As the radius is enlarged by a factor of $1/(1-\dot{\theta}_p^2)$, the resulting enclosed area becomes,
	\begin{equation} \label{ClassicLimit}
	A' = R_p^2 \int_0^T \frac{\dot{\theta}_p}{(1-\dot{\theta}_p^2)^2} \textrm{d}t.
	\end{equation}
	When $\dot{\theta}_p=\pi/T$, the area $A'=\pi R_p^2/[1-(\pi/T)^2]^2$ is identical to the one used in Eq.~(4) of the main text. Note, that this result is largely independent of atomic interactions, as shown in Fig.~\ref{fig:phasestability} and also Fig.~4b,d,f.
	
\end{appendix}
\end{document}